\documentclass[12pt]{article}

\usepackage{scicite}

\usepackage{times}
\usepackage[utf8]{inputenc}
\DeclareUnicodeCharacter{FB01}{fi}
\usepackage{graphicx}
\usepackage{dcolumn}
\usepackage{bm}
\usepackage{xcolor}
\usepackage{amsmath}

\newenvironment{sciabstract}{%
\begin{quote} \bf }
{\end{quote}}

\title{An experimental test of the geodesic rule proposition for the non-cyclic geometric phase}

\author
{Zhifan Zhou,$^{1}$ Yair Margalit, $^{1}$ Samuel Moukouri, $^{1}$ Yigal Meir,$^{1\ast}$ Ron Folman$^{1}$    \\
\\
\normalsize{$^{1}$Department of Physics, Ben-Gurion University of the Negev, Beer-Sheva 84105, Israel}\\
\\
\normalsize{$^\ast$To whom correspondence should be addressed; E-mail:  ymeir@bgu.ac.il}
}

\date{}

\begin{document} 


\baselineskip24pt


\maketitle


\begin{sciabstract}
  The geometric phase due to the evolution of the Hamiltonian is a central concept in quantum physics, and may become advantageous for quantum technology.
In non-cyclic evolutions, a proposition relates the geometric phase to the area bounded by the phase-space trajectory and the shortest
geodesic connecting its end points. The experimental verification
of this geodesic rule proposition has remained elusive for more than three decades. Here, we report an unambiguous experimental confirmation  of the geodesic rule for a non-cyclic geometric phase by means of a spatial SU(2) matter-wave interferometer, demonstrating, with high precision, the predicted phase sign change and $\pi$ jumps. We show the connection between our results and the Pancharatnam phase. Finally, we point out that the geodesic rule can be applied to obtain the red-shift in general relativity, enabling a completely new quantum tool to measure gravity.
\end{sciabstract}

The geometric phase\,(GP), the phase acquired over the course of an evolution of the Hamiltonian in parameter space, is a central concept in classical and in quantum physics\cite{pancha, berry,herzberg-longuet, simon,wilczek-zee,aharonov-anandan,berry_1987,samuel-bhandari,berry_2}.
Originally, the GP was defined only for an evolution of a system in a closed trajectory
in phase space, but later it was generalized to non-cyclic evolutions\cite{samuel-bhandari,bhandari}.  For the case of a 2-level system, where the evolution of the system can be described by a trajectory on the Bloch or Poincar\'e spheres, it has been proposed\cite{samuel-bhandari,bhandari} that, using a natural definition of the phase\cite{pancha},  the GP is given by half the area enclosed by the trajectory and the geodesic connecting the initial and final points.  A dramatic outcome of the proposed geodesic rule is that this non-cyclic phase changes sign when the trajectory moves from the upper to the lower hemisphere, resulting in a $\pi$-phase jump when the trajectory is half the circumference of a circle\cite{samuel-bhandari,bhandari}. While the GP for a closed trajectory has been measured experimentally in several physical systems\cite{PhotonExp,NeutronExp,QubitExp, 2014HalfSpin, 2014QuanCircuit}, the experimental verification of the GP during non-cyclic evolution remained elusive\cite{wagh, bhandary_2}. Utilizing an ultra-cold atom spatial interferometer we test the geodesic rule and the predicted SU(2) phase sign change and $\pi$ jumps.

Berry's original work\cite{berry} addressed a quantum system undergoing a cyclic evolution under the action of
a time-dependent Hamiltonian. When the Hamiltonian returns to its initial value, the quantum state acquires an extra GP in addition to the dynamical phase. Interestingly, this concept has been 
generalized\cite{samuel-bhandari} to a non-cyclic evolution of the system, where the parameters of the Hamiltonian do not return to their initial values. In addition to the fundamental interest in better understanding the non-cyclic behavior, it may also prove to be technologically advantageous. For example, as the system does not need to return to its original state, geometric operations may be done faster, e.g. geometric quantum gates\cite{Qubit}. In addition, metrology may be made more sensitive due to the expected phase sign change and phase jumps, e.g. in measuring a gravitational potential\cite{Wineland}.

\begin{figure}
\begin{center}
  \includegraphics[width=17.5cm, height=6.8cm]{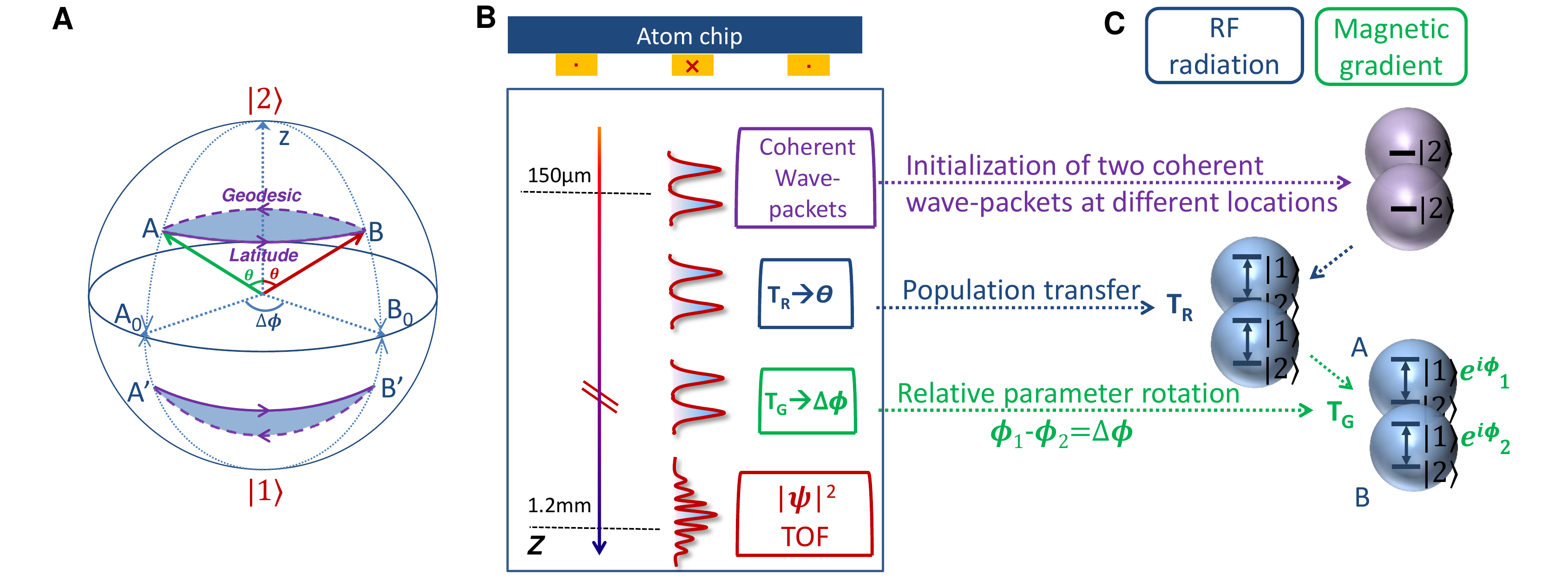}
\end{center}
\caption{{\bf (A)} An illustration of the geodesic rule\cite{samuel-bhandari, bhandari} on the Bloch sphere representing the 2-dimensional space defined by our physical 2-level system.
The green and red arrows represent the internal states A and B of the two spatially separated wave packets, $\Psi_A$ and $\Psi_B$ [see Eq.\,(1)].
The rotation angle from the north pole $\theta$ and the rotation $\Delta\phi$ along the latitude (continuous purple) represent the SU(2) operations applied in the experiment, where the former requires a radio-frequency (RF) pulse while the latter requires a magnetic gradient.
When $\theta=\pi/2$ the arrows lie on the equator of the Bloch sphere\,(A$_0$ and B$_0$).
The dashed purple curve is the geodesic joining
the points A\,(A') and B\,(B').  The GP is equal to one half of the blue area enclosed by the latitude and geodesic. The area's orientation\,(indicated by the arrows) is determined by the
geodesic rule. It is negative, counter clock-wise (northern hemisphere) and positive, clock-wise (southern hemisphere).
{\bf (B)} Experimental sequence\,(not to scale) of the longitudinal interferometer. The experiment is done in free fall. The final interference pattern (from which the total phase is obtained) develops after time-of-flight (TOF) free evolution in which the two wave packets expand and overlap. The pattern is then recorded by a CCD camera.
{\bf (C)} The evolution of the states during the sequence. After the preparation of two coherent wave packets at different locations, a RF pulse of duration $T_R$ is applied to manipulate $\theta$ and a magnetic field
gradient of duration $T_G$ is applied to manipulate $\Delta\phi$.}
\label{fig-sphere}
\end{figure}

The geometric interpretation of this non-cyclic GP takes an illuminative form for a 2-level system whose state can be described by two angles, $\Psi = (\cos \frac{\theta}{2} |2 \rangle + \exp\left(i\phi\right)\sin \frac {\theta}{2} |1 \rangle)$, which define a point on the Poincar\'e or Bloch spheres. The propagation of a state under a non-cyclic evolution of the Hamiltonian, from $\Psi_A$ to $\Psi_B$, characterized respectively by $\{\theta_A,\phi_A\}$ and  $\{\theta_B,\phi_B\}$, is represented by a curve connecting points $A$ and $B$ on the sphere. Using a natural definition of the phase\cite{pancha} -- where the relative phase between two arbitrary states is zero when the visibility of their interference pattern is maximal --  the GP associated with this propagation is determined by the geodesic rule:
it is given by half the area on the sphere bordered by the evolution curve
and the shortest geodesic connecting  $A$ and $B$. An illustration of the geodesic rule on the
Bloch sphere is shown in Fig.\,\ref{fig-sphere}, where $A$ evolves towards $B$, along the latitude of fixed $\theta_A=\theta_B=\theta$, and $\phi$ changes from $\phi_A$ to $\phi_B=\phi_A+\Delta\phi$ (the curve ${\cal C}_{AB}$).
The area corresponding to the GP, blue shaded in the figure, is enclosed by ${\cal C}_{AB}$ and by the  geodesic curve ${\cal G}_{AB}$ joining the points $A$ and $B$.
 If ${\cal C}_{AB}$ is on the northern hemisphere, ${\cal G}_{AB}$ is above\,(towards the north pole) ${\cal C}_{AB}$. But if ${\cal C}_{AB}$ is on the southern hemisphere, ${\cal G}_{AB}$ is below ${\cal C}_{AB}$, leading to a sign change of the GP as ${\cal C}_{AB}$ crosses the equator.

 Since the introduction of the geodesic rule, several studies have
tried to verify it experimentally with neutron\cite{wagh,filipp},
and atom\cite{AI2012,AI2016} interferometers. However, none of these studies
have unambiguously shown the two fundamental manifestations of the geodesic rule,
namely the $\pi$-phase jump and the sign change at the equator.
A limitation  common to these studies which precludes a clear conclusion is
the artificial change of the phase reference between the northern and
southern hemispheres (see for example criticism in Ref.\cite{bhandary_2}).
Phase jumps were reported in Ref.\cite{AI2009} but they were attributed to the negative sign of the transition amplitude
between hyperfine states, and the underlying physics behind the phase jump and its connection to the
geodesic rule was not discussed.

In this work we propose and realize a matter-wave experimental study using cold-atom spatial
interferometry \cite{Interf1,Interf2}. The advantages of our approach are that we use a spatial interference pattern to determine the phase in a single experimental run, we use a common phase reference for both hemispheres, and the relative phase is obtained by allowing $\Psi_A$ and $\Psi_B$ to expand in free flight and overlap, in contrast to previous atom-interferometry studies which required, for obtaining interference, an additional manipulation of the SU(2) parameters $\theta$ and $\Delta\phi$.
As a result we unambiguously confirm the geodesic rule for non-cyclic evolutions including the predicted sign change, and precisely confirm the predicted $SU(2)$ phase jumps.

Our full experimental procedure is detailed elsewhere\cite{FGBS,margalit, ClockComplementarity} and in the supplementary material\cite{SM} (see Fig.\,\ref{fig-expscheme}). The relevant part for the determination of the GP is sketched in Fig.\,\ref{fig-sphere}.
The $^{87}$Rb atom can be in either state $|1\rangle\equiv|F=2,m_F=1\rangle$ or $|2 \rangle\equiv|F=2,m_F=2\rangle$, where $F$ is the total angular momentum and $m_F$ is the projection. We start by preparing two atom
wave packets at different positions, both in an internal state $|2 \rangle$.
We first apply a uniform radio-frequency (RF) pulse, of time duration $T_R$, which transfers population from the $|2 \rangle$ state to $|1 \rangle$, shifting both wave packets from the north pole of the Bloch sphere to a point whose latitude $\theta$ depends on $T_R$ (Fig.\,\ref{fig-sphere}A). We then apply a magnetic-field gradient pulse of duration $T_G$, which results, due to the different magnetic moments of states $|1\rangle$ and $|2 \rangle$,  in a phase difference between these states, rotating both superpositions along a constant latitude on the Bloch sphere. Because of the difference in positions,
each wave packet experiences a different magnetic field, and thus will rotate by a different angle, ending up at points $A$ and $B$ in Fig.\,\ref{fig-sphere}A. The two states, after the application of both $T_R$ and $T_G$ can thus be written as
\begin{equation}
\Psi_A =\psi_A(r)  (\cos \frac{\theta}{2} |2 \rangle + \sin \frac {\theta}{2} |1 \rangle),~~\\
\Psi_B =\psi_B(r)  (\cos \frac{\theta}{2} |2 \rangle + \exp(i\Delta\phi)\sin \frac {\theta}{2} |1 \rangle),
\label{EQ-states}
\end{equation}
where $\theta$ is proportional to $T_R$, and $\Delta\phi$ to $T_G$.
 $\psi_A(r) $ and $\psi_B$ are the spatial components of the
respective states. There may also be an additional global
phase, identical for both $\Psi_A$ and $\Psi_B$, which plays no role in the interference between $\Psi_A$ and $\Psi_B$. To measure this interference, we allow enough time of flight for the two wave packets to free fall, expand, and overlap, before taking a picture using a CCD camera.

Fig.\,2 depicts the averaged interference
patterns\,(raw data CCD images) averaged over all values of $\theta$ in the upper (B) or lower (C) hemispheres, for $T_G=17\,\mu$s ($\Delta\phi\simeq\pi$). The value of $\theta$ was independently deduced from the relative populations of states $|1 \rangle$ and $|2 \rangle$ which are given by $\cos^2(\theta/2)$ and by $\sin^2(\theta/2)$, respectively (Figs.\,2A and S2).  The high visibility in both images indicates the existence of ``phase rigidity'', namely that the measured phase is independent of $\theta$ in each hemisphere. Moreover, the two data sets have a phase difference of $\pi$, which can also be deduced from the vanishing visibility in  Fig.\,2D, where the two data sets in (B) and (C) are joined.  Evidently, there is a sharp jump in the phase of the interference pattern as $\theta$ crosses the equator.

According to Eq.\,(\ref{EQ-states}), the interference phase $\Phi$, for general $\theta$ and $\Delta\phi$, is given by
\begin{equation}
\Phi=\arg\ \langle \Psi_A | \Psi_B \rangle = \phi_0 + \arctan \left \{ \frac{\sin^2(\theta/2)\sin\Delta\phi}{\cos^2(\theta/2)+\sin^2(\theta/2)\cos\Delta\phi} \right \},
\label{EQ-Phi}
\end{equation}
where $\phi_0= \arg \langle \psi_A(r) | \psi_B(r) \rangle $ is the phase associated with the evolution of the external degrees-of-freedom of the system\,(see S4 in \cite{SM}). Fig.\,3 depicts the interference phase, deduced from the raw data, as a function of $T_R$ for different values of $T_G$. The dashed lines in this figure are a fit to Eq.\,(2), with the fitting parameters $\phi_0$ (an overall vertical shift) and $\Delta\phi$. The excellent fit to the data allows us to determine with high precision the values of $\Delta\phi$ (Fig.\,3E).

The total phase\,(interference phase) $\Phi$ is a sum of two contributions, the geometric phase $\Phi_G$ and the dynamical phase $\Phi_D$. While both  $\Phi$  and $\Phi_D$ are gauge dependent, $\Phi_G=\Phi-\Phi_D$ is gauge independent \cite{mukunda-simon,dezela}. Substituting for the dynamical phase\cite{aharonov-anandan,bhandari,mukunda-simon,dezela},
we obtain\,(see S4 in \cite{SM})
\begin{equation}
\Phi_G=\arctan \left \{ \frac{\sin^2(\theta/2)\sin\Delta\phi}{\cos^2(\theta/2)+\sin^2(\theta/2)\cos\Delta\phi} \right \}-\frac{\Delta\phi}{2}(1-\cos\theta),
  \label{EQ_phase}
\end{equation}
where the gauge-dependent phase $\phi_0$ has dropped out.

\begin{figure}[htp!]
\begin{center}
  \includegraphics[width=17 cm, height=11cm]{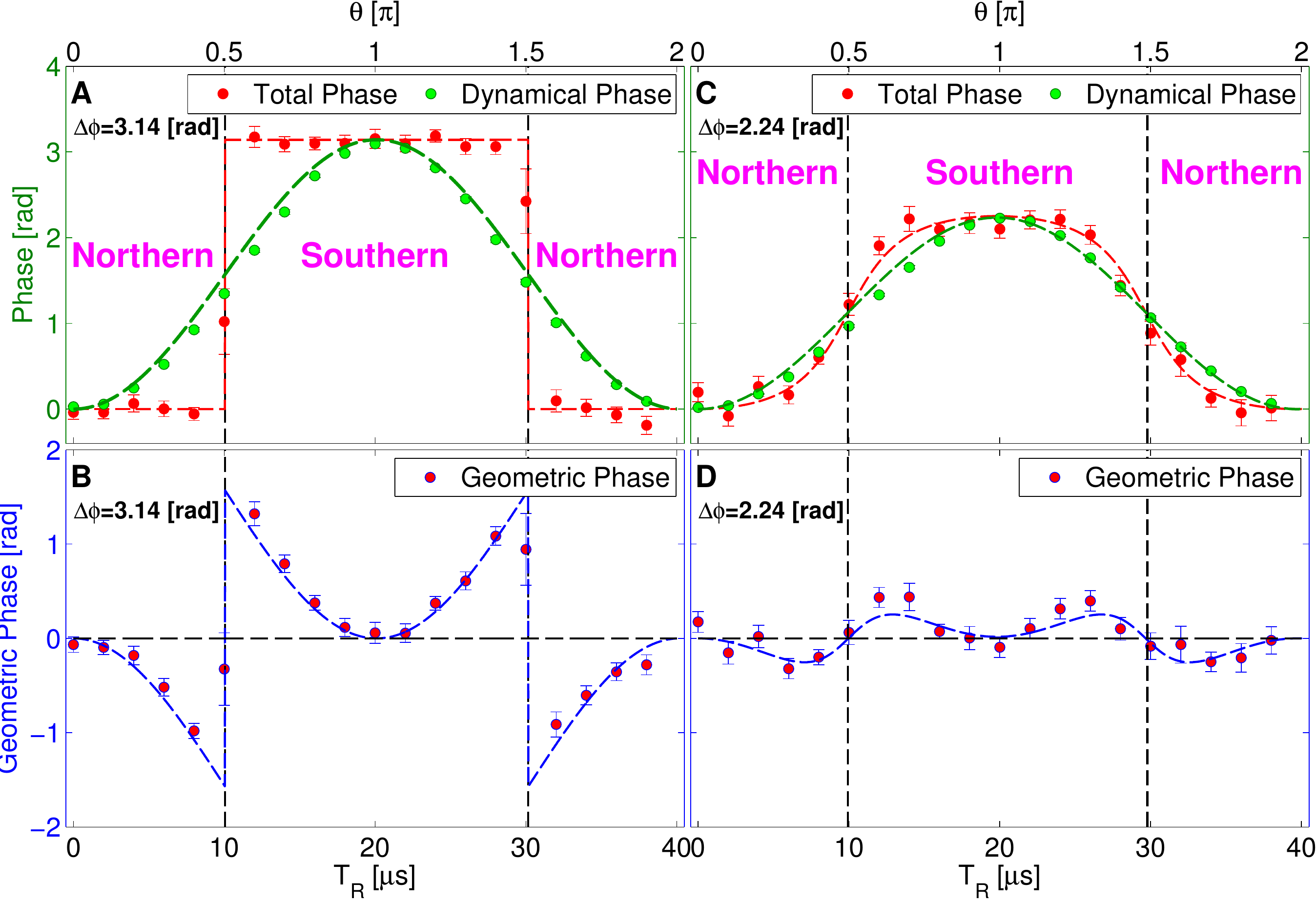}
\end{center}
\caption{Geometric SU(2) phase jump and sign flip, experiment\,(dots) versus theory\,(Eq.~\ref{EQ_phase}, dashed lines): {\bf (A)} Total phase and dynamical phase for $\Delta\phi=\pi$ as a function of $T_R$\,($\theta$). The total phase is directly measured from the imaged interference pattern\,(Fig.\,3), and the dynamical phase
$\frac{\Delta\phi}{2}(1-\cos\theta)$ is deduced from the independently measured values of $\theta$ and $\Delta\phi$.
{\bf (B)} The geometric phase $\Phi_G$ determined as the difference between the two sets of points appearing in ({\bf A}). The predicted sign change as the latitude crosses the equator is clearly visible.
The evident phase jump is due to the geodesic rule. When $\Delta\phi=\pi$ the geodesic must go through the Bloch sphere pole for any $\theta\ne\pi/2$. As the latitude approaches the equator (i.e. increasing $\theta$), the blue area in Fig.\,1 (twice $\Phi_G$) continuously grows to reach a maximum of $\pi$ in the limit of $\theta=\pi/2$. As the latitude crosses the equator, the geodesic jumps from one pole to the other pole, resulting in an instantaneous change of sign of this large area and a phase jump of $\pi$.
This plot exactly confirms the prediction in \cite{bhandari}.
{\bf (C)} Total phase and dynamical phase for $\Delta\phi=2.24$\,rad.
{\bf (D)} $\Phi_G$, determined as the difference between the two sets of points appearing in ({\bf C}). The predicted sign change is again visible. However, in the case of $\Delta\phi=2.24$\,rad the geodesic line does not go through the pole and as the latitude approaches the equator $\Phi_G$ actually reduces\,(after reaching its maximum for an intermediate $\theta$), so no abrupt phase jump is expected.
  }
\label{fig-GeoPhase}
\end{figure}

Fig.\,\ref{fig-GeoPhase} displays $\Phi$, $\Phi_D$ and the resulting $\Phi_G$, for two values of $\Delta\phi$, where the first term on the RHS of Eq.\,(\ref{EQ_phase}) is given by $\Phi$, the phase of the interference pattern, while the second is evaluated for the experimentally determined values of $\theta$ and $\Delta\phi$.
The dashed lines in panels (B) and (D) correspond to the geodesic rule - half the area between the geodesic and the trajectory, with the correct sign.
A very good agreement between data and the theoretical predictions is observed.
This constitutes a complete verification of the hitherto elusive geometric phase associated with non-cyclic evolution in an SU(2) system, and confirms unambiguously the long standing predictions, including a precise observation of the geodesic rule, the phase sign change, and the $\pi$ phase jump.

Finally, we make a fundamental connection between our experiment and the Pancharatnam phase\cite{pancha}.
We begin by noting that in the case $\phi_0=0$ we have $\arg \langle \Psi_A  | 2\rangle=0$
and  $\arg \langle 2 | \Psi_B \rangle=0$, and then the states $A$, $|2 \rangle$
and $B$ fulfill the Pancharatnam consecutive in-phase criterion \cite{pancha, Dijk2010}. It then follows that
$\arg \langle A | B \rangle$ is given by half the area $\sigma$ of the
spherical triangle defined by these three states on the Bloch sphere, namely, the area in between three geodesic lines.
The area $\sigma$ of the spherical triangle defined by the two arcs joining the
north pole and the points A and B respectively is given by the relation
$\tan (\sigma/2)=\tan^2(\theta/2) \sin(\phi)/[1+\tan^2 (\theta/2) \cos \phi]$, which is
identical to Eq.\,(\ref{EQ-Phi}) with $\Phi=\sigma/2$ (for $\phi_0=0$). This geometric interpretation
of $\Phi$ yields an explanation of the observed phase rigidity for $\Delta\phi=\pi$:
When the two points are in the northern hemisphere,
the geodesic between the two points goes through the north pole.
The enclosed area is zero, hence $\Phi=0$. When the two points are in the
southern hemisphere, the geodesic goes through the south pole, with an area of $2\pi$, resulting in a jump of $\pi$ in the value of $\Phi$ (Fig.\,2).
The geometric interpretation of our experiment is now evident, namely, what is measured in the experiment (the interference-pattern phase) is the Pancharatnam phase $\Phi_P$.
The difference between the areas associated with $\Phi_P$ and $\Phi_D$ gives the light blue area in Fig.\,1, associated with $\Phi_G$. This now naturally explains both the sign change of $\Phi_G$ as the latitude crosses the equator, as well as the phase jump for $\Delta\phi=\pi$\,(Fig.\,4).

As an outlook we consider a situation in which the two wave packets are viewed as a split wave packet of a single clock, where $\theta=\pi/2$ for a perfect 2-level clock\cite{margalit,ClockComplementarity}.
When we place the two wave packets along a vertical line parallel to gravity at different distances from earth, they are exposed to different proper times.
In the experiment described in this paper, the relative phase $\Delta\phi=\Delta(E_1-E_2)\times t/\hbar$ between the wave packets is determined by a magnetic gradient which changes the energy splitting $E_1-E_2$ between states $|1\rangle$ and $|2\rangle$ (i.e. $\Delta(E_1-E_2)$ is the difference of energy splitting between two wave packets $\Psi_A$ and $\Psi_B$), while time\,(from the moment the two wave packets were allowed to free fall) is the same for both wave packets.
However, the same GP situation occurs when the magnetic gradient is zero and consequently the splitting $E_1-E_2$ is identical for the two wave packets, but time elapsed is different for the two wave packets due to the different red-shift\,(with time difference $\Delta t$). In this case, we have $\Delta\phi=(E_1-E_2)\times \Delta t/\hbar$ and the same theory presented in this paper
may be used to analyze via the GP an experimental situation on the interface between quantum mechanics and general relativity. Moreover, by scanning $\theta$ around
$\pi/2$\,(i.e. change the relative populations of the  $|1\rangle$ and $|2\rangle$ states from below to above half) one should observe a sign change which may allow for the construction of a novel type of gravitational sensor.

\section*{Acknowledgments}
This work is funded in part by the Israel Science Foundation (grants no.\,292/15 and 856/18), the German-Israeli DIP project (Hybrid devices: FO 703/2-1) supported by the DFG, and the European ROCIT consortium for stable optical clocks. We thank the BGU nano-fabrication facility for the high quality chip.
 We also acknowledge support from the Israeli Council for Higher Education and from the
 Ministry of Immigrant Absorption (Israel).

\begin{figure}[htp!]
\begin{center}
  \includegraphics[width=14 cm, height=10.5cm]{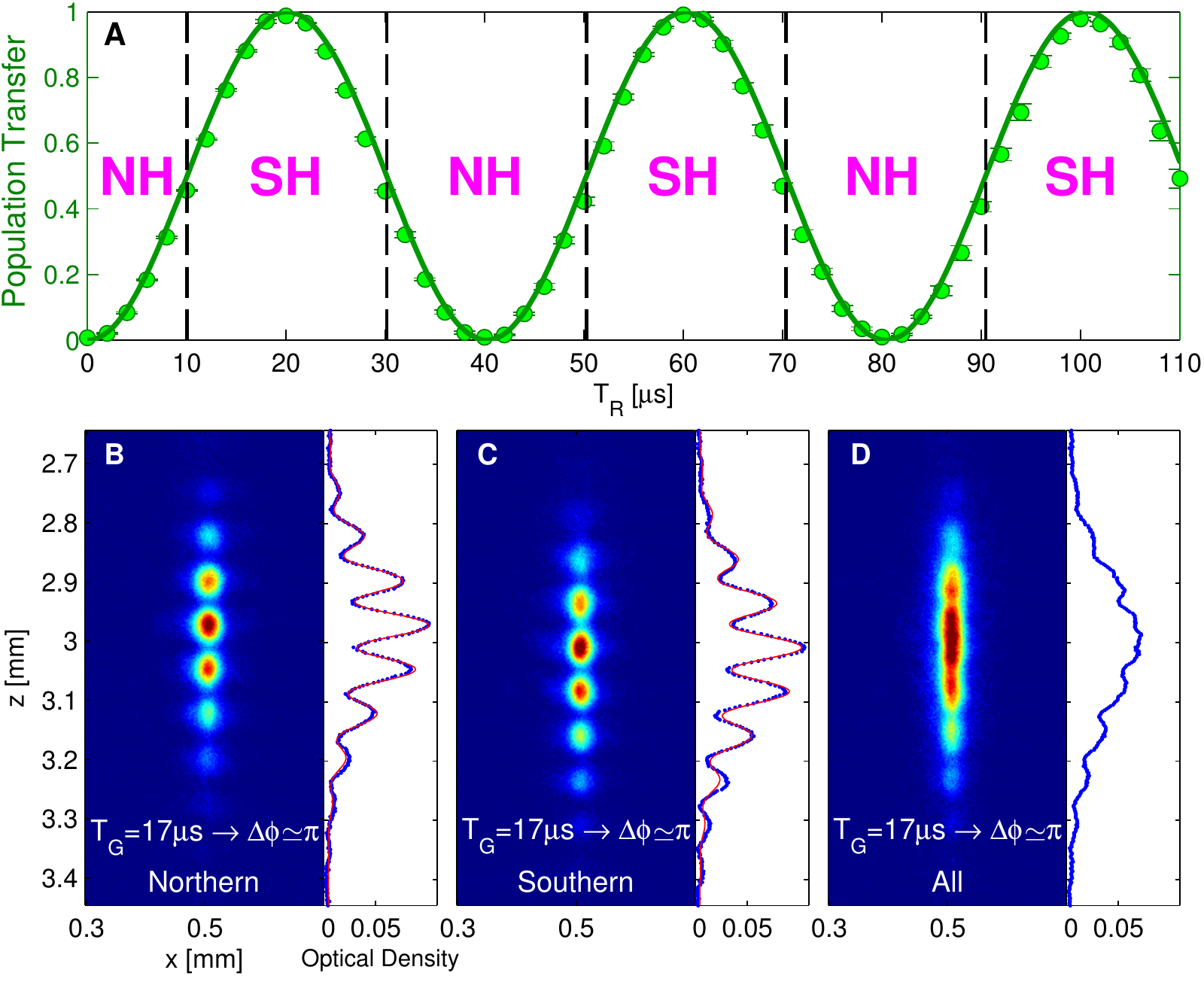}
\end{center}
\caption{Experimental $\pi$ phase jump: {\bf (A)} Population transfer to state $|1\rangle$ versus the duration of the RF radiation pulse $T_R$, for which 20$\,\mu$s correspond to total population transfer ($\theta=\pi$ in Fig.\,1). With this independent measurement we determine $\theta$ for our SU(2) operations.
{\bf (B)} The averaged CCD image of interference when the Bloch vectors are all in the northern hemisphere [NH data points specified in (A)], with $\Delta\phi\simeq\pi$. The high visibility indicates the existence of ``phase rigidity'', namely that the phase is independent of $\theta$.
The phase returned by the fit is 1.13 $\pm$0.02 relative to a  fixed  reference point, and the visibility is 0.55 $\pm$ 0.01\,(see \cite{SM} for definition).
{\bf (C)} The averaged picture of the second half of the data, in which the Bloch vectors are all pointing in the southern
hemisphere [SH data points specified in (A)], with $\Delta\phi\simeq\pi$. A phase jump is clearly visible. The phase is 4.34 $\pm$ 0.03 relative to the fixed reference point which is common to both pictures, and the visibility is
0.52 $\pm$ 0.01. The phase difference between (B) and (C) is thus 3.21 $\pm$ 0.05, close to $\pi$. The data included in these images\,(in total about 330 consecutive experimental shots without post-selection or post-correction) is presented in Fig.\,3B.
{\bf (D)} The averaged picture of all the data for $\Delta\phi\simeq\pi$. The
visibility is 0.03 $\pm$ 0.01. The low visibility clearly shows that the phase jump has a value close to $\pi$. Single-shot data is presented in Fig.\,3B and single-shot images are presented in Fig.\,S3 in \cite{SM}.}
\label{fig-rawfigure}
\end{figure}

\begin{figure}[htp!]
\begin{center}
  \includegraphics[width=15 cm, height=12.cm]{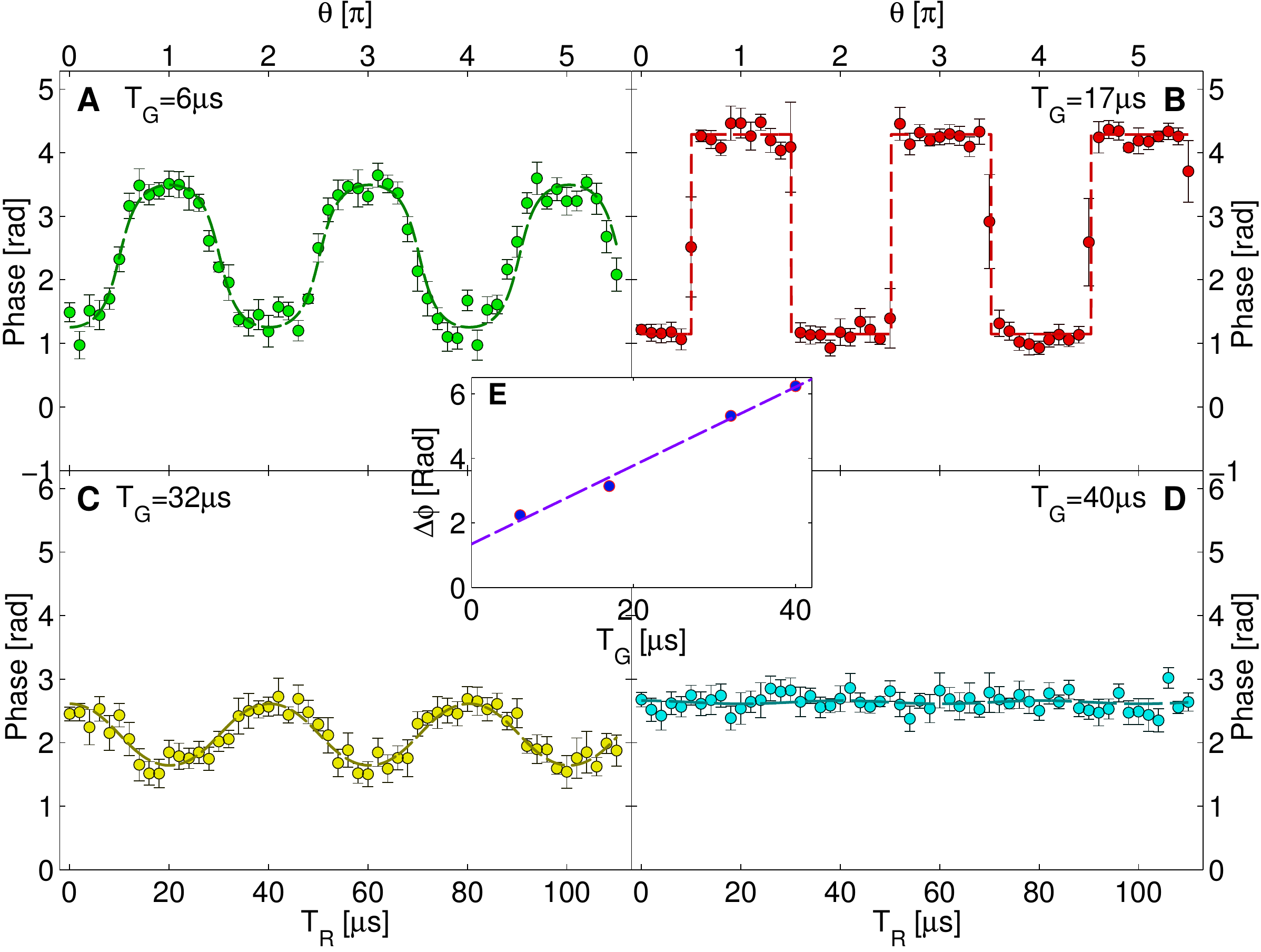}
\end{center}
\caption{Interference-pattern phase: {\bf (A-D)} Total phase $\Phi$ as a function of $T_R$\,($\theta$) for $T_G$ equal 6, 17, 32 and 40\,$\mu$s, respectively.
Each data point is an average of 6 experimental cycles\,(errors are Standard Error Mean). The dashed lines are a fit to Eq.\,(2), which allows us to determine $\Delta\phi$ for our SU(2) operations.
The fit returns the values $\Delta\phi=2.24$\,({\bf A}), $\Delta\phi=3.14$\,({\bf B}), $\Delta\phi=5.31\equiv2\pi-0.97$\,({\bf C}) and $\Delta\phi=6.23\equiv2\pi-0.05$\,({\bf D}) radians, respectively\,(manifested in the graph as the peak-to-valley amplitude if we consider the periodicity of 2$\pi$ when defining a phase). The fit also returns a base-line phase $\phi_0$. Finally, the phase rigidity and the phase jump observed in Fig.\,2 are clearly visible in ({\bf B}). {\bf (E)} The linear mapping from $T_G$ to $\Delta\phi$. As seen in the graph ($T_G=0$) we have a fixed background gradient equivalent to $\Delta\phi=1.35$. }
\label{fig-rawexp}
\end{figure}

\newpage

\pagebreak

\pagebreak

\pagebreak

\bigskip
\bigskip
\bigskip

\newpage
\newpage
    
\section*{Supplementary materials}
Materials and Methods\\
Supplementary Text\\
Figs. S1 to S4

\clearpage
\renewcommand{\theequation}{S\arabic{equation}}
\renewcommand{\thefigure}{S\arabic{figure}}
\setcounter{figure}{0}
\setcounter{equation}{0}

{\bf \it S.1 Detailed experimental scheme}

\begin{figure}
\begin{center}
  \includegraphics[width=15 cm, height=12.cm]{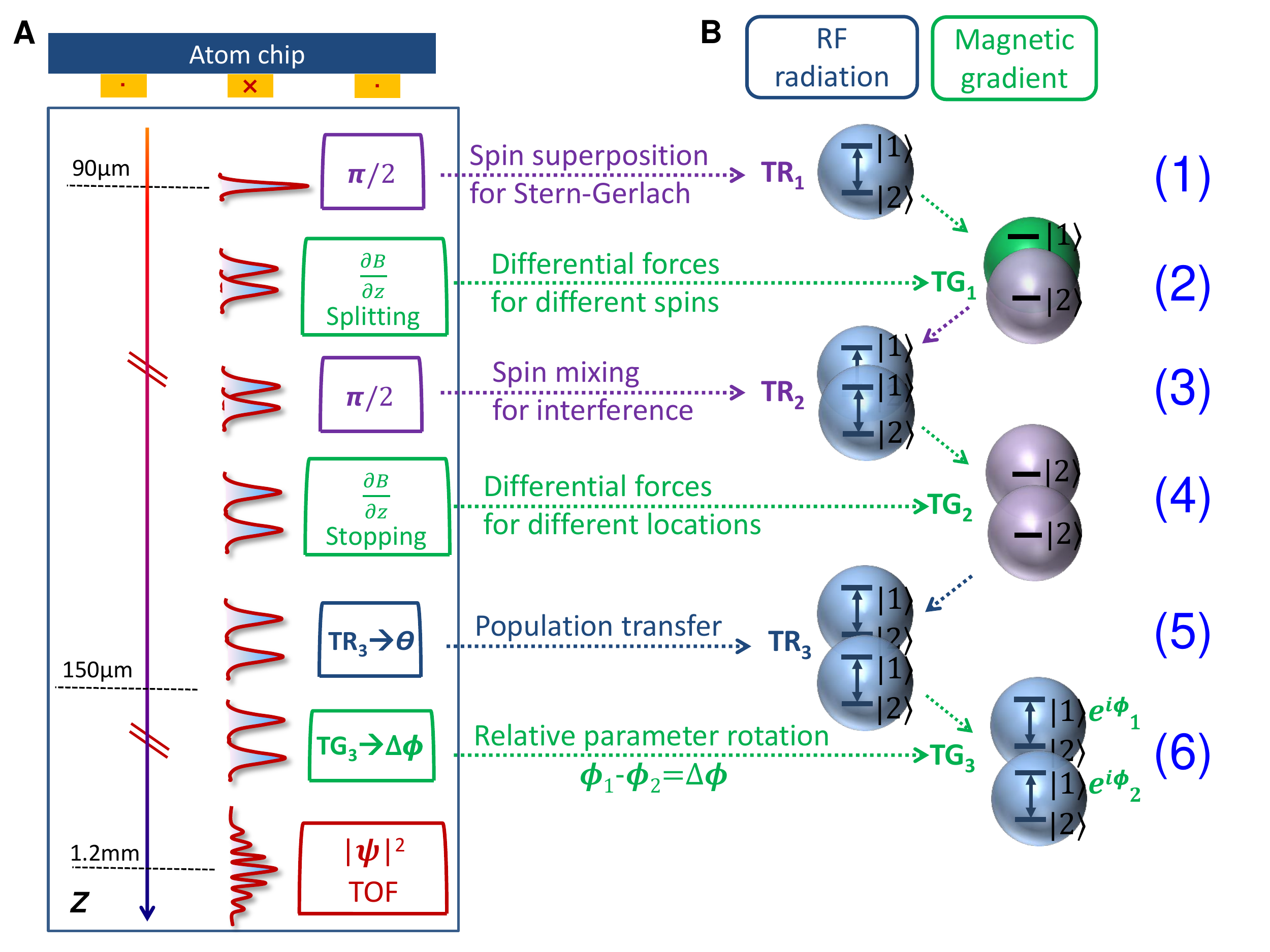}
\end{center}
\caption{ {\bf (A)} Detailed sequence\,(not to scale) of the longitudinal interferometer along the z axis (the direction of gravity).
{\bf (B)} The evolution of the states along the sequence.
After the atoms are released from the trap, one radio-frequency $\pi/2$ pulse\,($TR_1$) is applied
to create an equal superposition of $|2 \rangle\equiv|F=2,m_F=2\rangle$ and $|1 \rangle\equiv|F=2,m_F=1\rangle$ states\,(1). These two spin states are then exposed to a differential force created by a magnetic gradient pulse ${\partial B}/{\partial z}$
of duration $TG_1$\,(2),
generated by currents in the atom chip wires, leading to different accelerations, and, as a result, different positions and different final velocities of the two states.
A second $\pi/2$ pulse\,($TR_2$)\,(3) is applied to mix the spins in each one of the wave packets and then, to stop the relative velocity of the wave packets, a second
second magnetic gradient pulse\,($TG_2$)\,(4) is applied to yield differential forces for the same-spin states which are at different locations.
As during $TG_2$, the $|1 \rangle$ state from the two wave packets are pushed outside the experimental zone, the system then consists of two wave packets in the $|2 \rangle$ state\,
(separated along the z axis, with zero relative velocity).
This 2-level system is initialized with
a third RF pulse\,(5) of duration $TR_3$\,($T_R$ in the text), after which the relative phase of the
two wave packets ($\Delta\phi$) may be changed by applying the third magnetic field
gradient\,(6) of duration $TG_3$ ($T_G$ in the text). Last,
before an image is taken, the wave packets are allowed to expand and overlap.}
\label{fig-expscheme}
\end{figure}

The experiment is realized in an atom chip set-up\,[M. Keil, O. Amit, S. Zhou, D. Groswasser, Y. Japha, and R. Folman, 'Fifteen years of cold matter on the atom chip: promise, realizations, and prospects', J. Mod. Opt. {\bf 63}, 18 (2016)].  We present the detailed experimental scheme in Fig.\,\ref{fig-expscheme} which includes the 2-level system preparation. We first prepare a Bose-Einstein condensate (BEC) of about $10^4$ $^{87}$Rb
atoms in the state $|2 \rangle\equiv|F=2,m_F=2\rangle$ in a magnetic trap located 90\,$\mu$m below the chip surface. After the BEC atoms are released from the trap, the entire experimental sequence takes place in the presence of a homogeneous magnetic bias field of 36.7\,Gauss in the y direction\,(z is the direction of gravity), which creates an effective 2-level system\,(with $|1 \rangle\equiv|F=2,m_F=1\rangle$) via the non-linear Zeeman
effect with $E_{ij}=E_{21}\approx h\times25\,$MHz (where i,j are the $m_F$ numbers, all in the $F=2$ manifold), and $E_{21}-E_{10}\approx h\times180\,$kHz. We then apply a radio-frequency\,(RF) pulse\,(duration $TR_1$, where typically $10\,\mu$s give rise to a $\theta=\pi/2$ rotation) to prepare a spin superposition $(|1\rangle+|2\rangle)/\sqrt{2}$ between the
$|2\rangle$ and $|1 \rangle$ states. A magnetic gradient pulse ${\partial B}/{\partial z}$ of duration $TG_1=4\,\mu$s, generated by currents in the atom-chip wires, is applied to create the Stern-Gerlach splitting, in which the different spins are exposed to differential forces. In order to enable interference between the two wave packets ($|2\rangle$ and $|1\rangle$ are orthogonal), a second $\pi/2$ pulse\,($TR_2$) is applied to mix the spins. To stop the relative velocity of the wave packets a second magnetic gradient pulse\,($TG_2$) is applied to yield differential forces for the same-spin states at different locations. A spatial superposition of two wave packets in state $|2 \rangle$ now exists\,(separated
along the z axis, with zero relative velocity). Note that during $TG_2$, the $|1 \rangle$ state from the two wave packets are pushed outside the experimental zone. The control of $\theta$ introduced in Fig.\,1A is realized by a third RF pulse of duration $TR_3$ ($T_R$ in the main text). The relative rotation between the two wave packets $\Delta\phi$ may be changed by applying a third magnetic field gradient of duration $TG_3$ ($T_G$ in the main text). The wave packets are then allowed to expand (during time-of-flight of $\sim$10\,ms, much larger than the reciprocal of the trap frequency $\sim$500\,Hz) and overlap to form the interference pattern. An image based on the
absorption-imaging is taken in the end (Fig.\,S3).

The magnetic gradient pulses are generated by three parallel gold wires located on the chip surface with 10-mm length, 40-$\mu$m width and 2-$\mu$m thickness.
The chip wire current is driven using a simple 12.5 V battery, and modulated using a home-made current shutter. The three parallel gold wires are separated by
100\,$\mu$m (center-to-center) and the same current runs through them in alternating directions. The benefit of using this 3-wire configuration instead of a single gold wire is that a 2D quadrupole field is created
at $z=100\,\mu$m below the atom chip. As the magnetic instability is proportional to the field strength, and as the main instability originates in the gradient pulses\,(the bias fields from external coils are very stable), positioning the atoms near the middle\,(zero) of
the quadrupole field significantly reduces the magnetic noise while maintaining the strength of the magnetic gradients.

\bigskip

{\bf \it S.2 Determination of the population transfer and the value of $\theta$}

\begin{figure}
\begin{center}
  \includegraphics[width=17 cm, height=11cm]{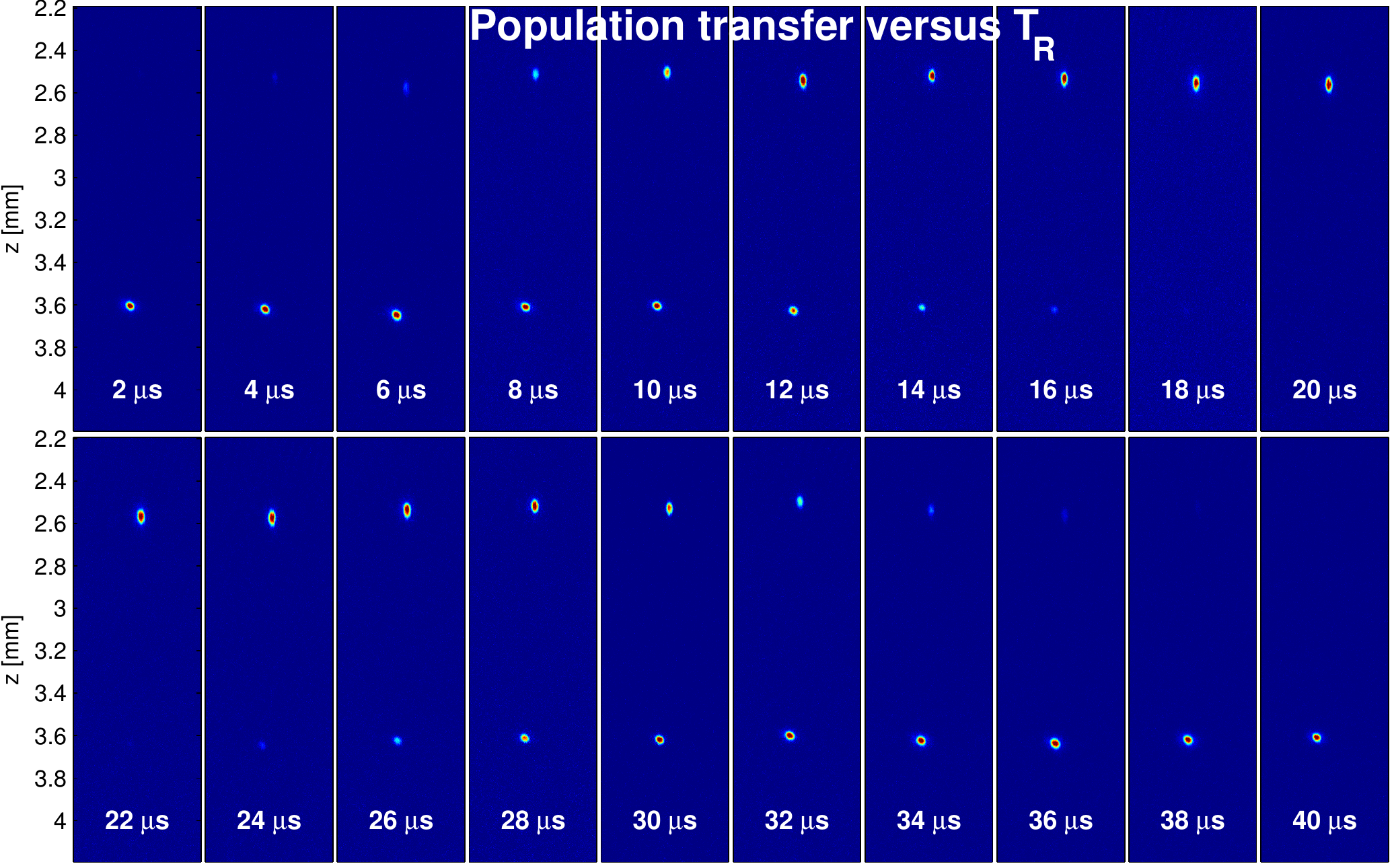}
\end{center}
\caption{The population transfer versus $T_R$ is measured in an independent experiment by applying a strong magnetic gradient after $T_R$. Due to the Stern-Gerlach effect, the $m_F=1$ and $m_F=2$ parts are shifted to different regions of space when the absorption imaging is done to evaluate the atom number respectively. The absorption imaging is based on the comparison between the intensity $I$
of a light pulse going through the atoms and the intensity $I_0$ of a reference light pulse that propagates in the absence of atoms and the Beer's law, $I(x_i,z_j)=I_0(x_i,z_j) e^{-OD(x_i,z_j)}$. The optical density\,(OD) is proportional to the column density of the atoms at a given position $\int n(x,y,z) dy$, where x and z are the object plane positions corresponding to $x_i$ and $z_j$, respectively. The number of atoms
$N(x_i,z_j)$ imaged by the pixel is $N(x_i, z_j)=\frac{A}{\sigma_0}OD(x_i,z_j)$, where $A$ is the pixel area in the object plane, $\sigma_0=3\lambda^2/2\pi$ is the cross-section for resonant atom-light scattering, and $\lambda\approx780$\,nm is the optical transition wavelength. The total atom number equals to $\int N(x,z) dxdz$.
We can then reliably determine the relation between population transfer and $T_R$ as presented in Fig.\,2A, e.g.\,10$\,\mu$s corresponds to $\theta=\pi$/2, 20$\,\mu$s corresponds to $\theta=\pi$ and 40$\,\mu$s corresponds to $\theta=2\pi$.  }
\label{fig-PopuVsTR}
\end{figure}

In Fig.\,\ref{fig-PopuVsTR} we explain how the values of $\theta$ are obtained from the measurement of population
transfer when we apply $TR_3$ ($T_R$ in the main text). Stern-Gerlach splitting is used to separate the $m_F=1$ and $m_F=2$ parts and absorption
imaging
is done to evaluate the atom-number respectively. See the details in the figure caption.

\bigskip

{\bf \it S.3 The CCD image of the interference pattern while $\theta$ is scanned}

In Fig.\,S3 we show the raw data of the interference patterns which are displayed in
Fig.\,2 (averaged over numerous values of $\theta$) and in Fig.\,3B (where the phase for different values of $\theta$ is presented), when $TG_3$ ($T_G$ in the main text) equals 17$\,\mu$s ($\Delta\phi\simeq\pi$).
The whole scanning range of $T_R$ is 40\,$\mu$s, corresponding to one full cycle\,(2$\pi$) of the Rabi oscillation.
The phase of the interference pattern is found to be rigid when the Bloch vector is located in the northern hemisphere or in the southern
hemisphere, with a $\pi$ phase-jump in between.

The interference pattern is fitted with the function $A$exp$[-\frac{(z-z_{CM})^2}{2\sigma_z^2}$]\{1+$v$ sin[$\frac{2\pi}{\lambda}$$(z-z_{ref}$)+$\Phi$]\}+c, where A is a constant related the optical density in the system, $z_{CM}$ is the center-of-mass\,(CM) position of the combined wave packet at the time of imaging, $\sigma_z$ is the Gaussian width of the combined wave packet obtained after time-of-flight, $\lambda=\frac{ht}{md}$ is the fringe periodicity, $v$ is the visibility, $z_{ref}$ is a fixed reference point, c is the background optical density from the absorption-imaging, and $\Phi$ is the phase of the interference pattern which appears in Eq.\,2. In the fringe periodicity $\lambda=\frac{ht}{md}$, $h$ is the Planck constant, $t$ is the duration of time-of-flight, $m$ is the mass of $^{87}$Rb atom, and $d$ is the distance between the two wave packets. In Fig.\,3 we measure the dependence of $\Phi$ on $\theta$\,($T_R$) for a fixed $T_G$ and then fit the data to Eq.\,(2), returning values for both $\phi_0$ and $\Delta\phi$.

\bigskip

{\bf \it S.4 Geometrical phases for different values of $\Delta\phi$}

Here we describe the approach used to derive the expression of $\Phi_G$ in
Eq.\,(\ref{EQ_phase}).  Mukunda and Simon developed a general
formalism called the quantum kinematic approach for the geometric phase in quantum systems\,[N. Mukunda and R. Simon, 'Quantum Kinematic Approach to
  the Geometric Phase. I. general Formalism', Annals of Physics {\bf 228},
  205 (1993)].

In the formalism of Mukunda and Simon, a one-parameter smooth curve is defined
from a vector $\psi$ belonging to an Hilbert space $\mathcal{H}$,
$\mathcal{C}=\{ \psi(s) \in \mathcal{N}_0, s \in [s_1,s_2] \}$.
$\mathcal{N}_0$ is the subset of unit vectors of $\mathcal{H}$. It is
important to note that the curve $\mathcal{C}$ is not necessarily closed.
The only requirements of the theory are the smoothness of $\mathcal{C}$, i.e,
$\psi(s)$ should be differentiable and the non-orthogonality of the initial
and final states. The geometric phase is given by
\begin{equation}
  \Phi_G=\Phi-\Phi_D,
\end{equation}
\noindent where $\Phi$ is the total phase. $\Phi_D$ is the dynamical phase arising from the energy
dependence on $s$ during the evolution. This general formalism naturally
reduces to the evolution under the time-dependent Schr$\ddot{o}$dinger
equation if the parameter $s$ is time. The curve $\mathcal{C}$ is the
trajectory of the wave-function during the propagation time $0 \le t \le T$.

The total phase $\Phi$ during an evolution along $\mathcal{C}$ is given by,
\begin{equation}
  \Phi=\arg \langle \psi(s_1)|\psi(s_2) \rangle.
\end{equation}

Taking $\psi(s) =  (\cos \frac{\theta}{2} |2 \rangle + \exp(is\Delta\phi)\sin \frac {\theta}{2} |1 \rangle)$, and $\left \{s_1,s_2\right \}=[0,1]$  we find for the total
phase
\begin{equation}
  \Phi=\arctan \left \{ \frac{\sin^2(\theta/2)\sin\Delta\phi}{\cos^2(\theta/2)+\sin^2(\theta/2)\cos\Delta\phi} \right \},
\end{equation}

\noindent where we should add to $\Phi$ the phase $\phi_0$ arising from the
evolution of the spatial part. This yields Eq.\,(\ref{EQ-Phi}) of the main
text.

\noindent The dynamical phase $\Phi_D$ can be calculated from the integral of the evolution
curve $\mathcal{C}$\cite{samuel-bhandari},
\begin{equation}
  \Phi_D=\mathrm{Im} \int_{s_1}^{s_2} \langle \psi(s)|{\dot \psi(s)} \rangle ds.
\end{equation}

  \noindent We find,
\begin{equation}
  \Phi_D=\frac{\Delta\phi}{2}(1-\cos\theta),
\end{equation}

to which the phase $\phi_0$ should also be added. Subtracting $\Phi_D$ from
$\Phi$ yields the expression for $\Phi_G$ of Eq.\,(\ref{EQ_phase}).
$\Phi_G$ is more suitable to use for analysis because gauge-dependent phases in
$\Phi$ and $\Phi_D$ mutually cancel.

In Fig.\,S4 we present the detailed theoretical behavior of $\Phi_G$ (Eq.\,3 of the main
text) as a function of $\theta$ and $\Delta \phi$. The characteristics of $\Phi_G$ are the singularity
at $\Delta \phi=\pi$ and $\theta=(n+1/2)\pi$\,(where $n$ is an integer), and the change of sign when $\theta$
goes across these values.  This result was originally obtained in [R. Bhandari, SU(2) phase jumps and geometric phases, Physics Letters A {\bf 157}, 221 (1991), Fig.\,4].

\begin{figure}
\begin{center}
  \includegraphics[width=17 cm, height=5.5cm]{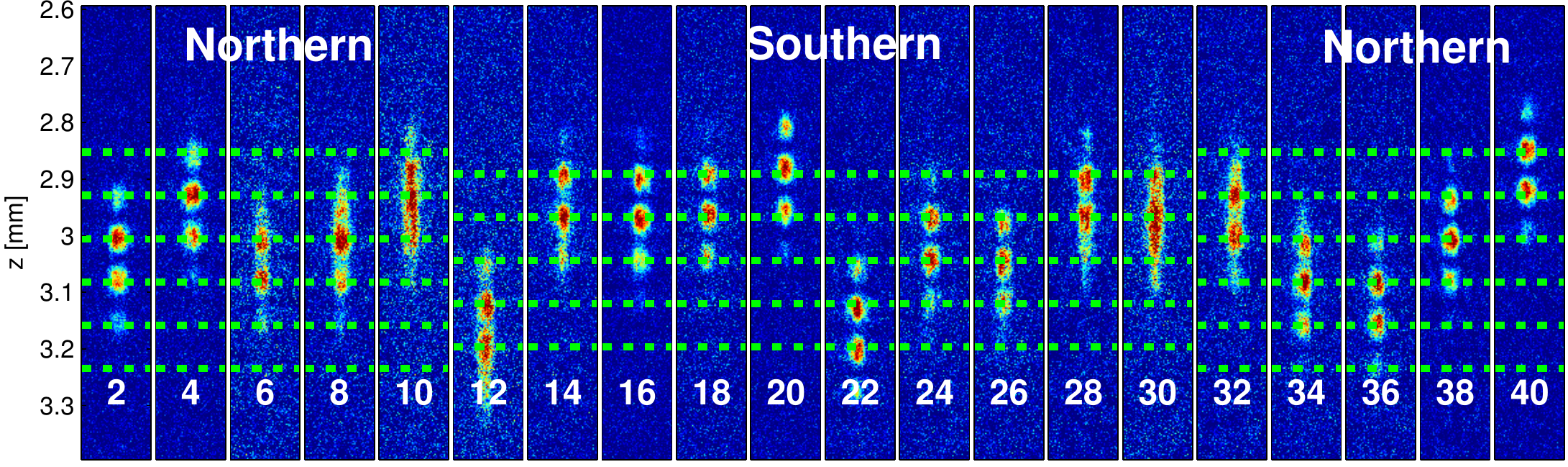}
\end{center}
\caption{The interference pattern versus $T_R$ when $T_G=17\,\mu$s ($\Delta\phi\simeq\pi$). The number in each sub-figure indicates the duration of $T_R$ in $\mu$s. When the Bloch vectors are in the northern hemisphere, the interference phase are seen to be rigid (fixed). When the Bloch vectors cross the equator at $T_R$=10$\,\mu$s, there is a $\pi$ phase jump. The interference phase will jump by another $\pi$ when the vectors cross the equator again at $T_R=30\,\mu$s. Namely, phase rigidity appears when the Bloch vectors are located in either the northern or the southern hemisphere, with a $\pi$ phase-jump in between, as presented in Fig.\,2B-D and Fig.\,3B. The fluctuations in the interference pattern's location are due to fluctuations in the initial conditions from shot-to-shot, while the inferred phase is stable, as explained in [S. Machluf, Y. Japha and R. Folman, Coherent Stern-Gerlach momentum splitting on an atom chip,
Nat. Commun. {\bf 4}, 2424 (2013)]. }
\label{fig-PatternVsTR}
\end{figure}

\begin{figure}
\begin{center}
  \includegraphics[width=12 cm, height=8.cm]{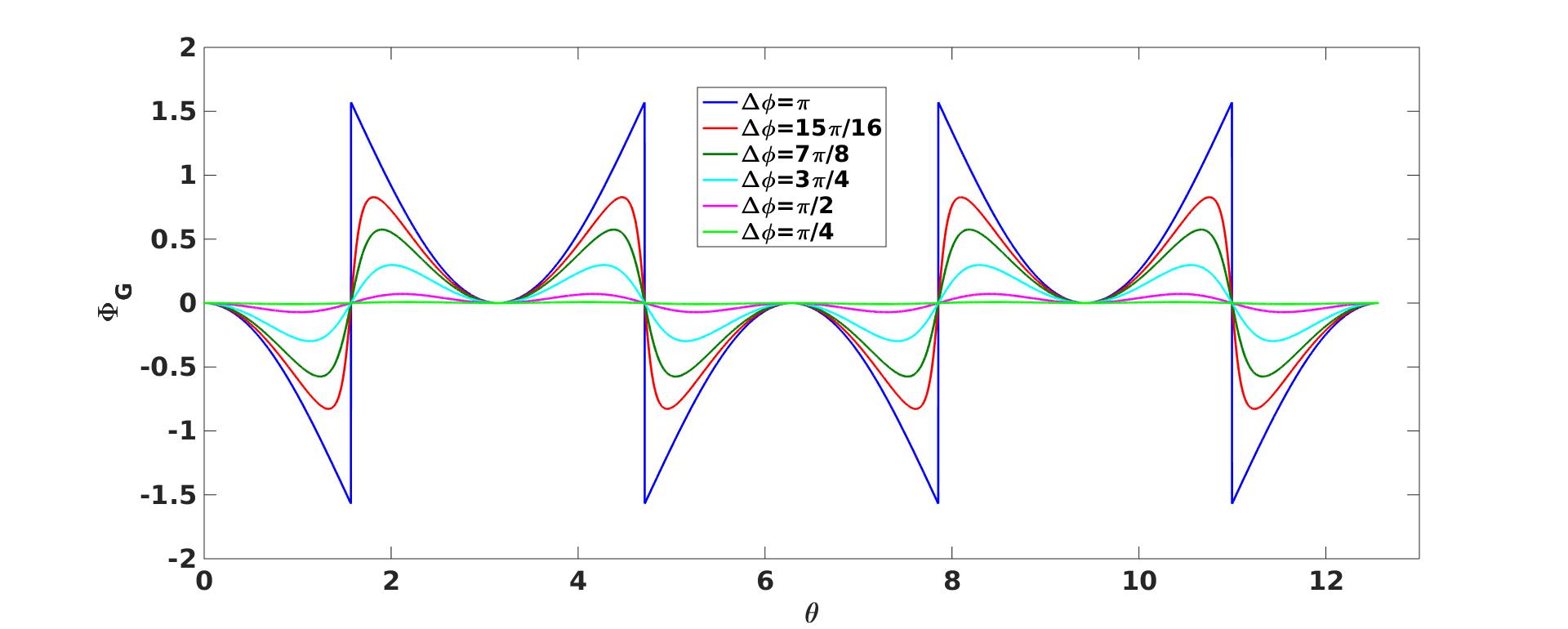}
\end{center}
\caption{The geometric phase $\Phi_G$ as given by Eq.\,3 as a function of $\theta$ for different values of $\Delta\phi$.}
\label{fig-geo}
\end{figure}

\end{document}